\documentclass[aps,prlb,twocolumn,groupedaddress,epsgraphics]{revtex4}

\usepackage{amsmath}
\usepackage{graphics}

\begin{document}

\title{Qubit entanglement in multimagnon states}

\author{J.S. Pratt}

\affiliation{Center for Quantum Information and
Department of Physics and Astronomy,
University of Rochester, Rochester, NY 14627}

\begin{abstract}
    The qubit entanglement induced by quasiparticle excitations in the
    Heisenberg spin chain and its relationship to the Bethe Ansatz
    structure of the eigenmodes is studied.  A phenomenon called
    entanglement quenching, which suppresses eigenstate entanglement,
    is described and shown to be mediated by Goldstone magnons. 
    Scattering states are characterized by short-range entanglement,
    and never exhibit entanglement at the longest range.  In contrast,
    bound states have complex long-range entanglement structures.
\end{abstract}

\pacs{PACS number(s):75.10.Jm,03.67.-a}

\maketitle

\section{Introduction}

The importance of quantum entanglement as a technological resource for
quantum computing \cite{ForcerHeyRossSmith} and quantum communication
\cite{BennettWiesner}, as an experimental requirement for Bell's
inequality violations \cite{OuMandel} and other foundational tests of
quantum mechanics, and as a central subject in the mathematical
analysis of highly correlated states \cite{Keyl} is now widely
recognized.  A fundamental theoretical problem is to understand the
complex entanglement structure naturally present in systems as diverse
as the Rindler vacuum \cite{PeresTerno}, the photons produced through
spontaneous parametric downconversion \cite{LawEberly}, and the
exchange-coupled spins of a magnet \cite{ArnesenBoseVedral}.  Much of
the work on this problem to date has focused on ground state
entanglement and its relation to quantum phase transitions \cite{QPT
refs}, or on thermal state entanglement and its modulation by magnetic
fields \cite{ArnesenBoseVedral}, \cite{thermal entanglement refs}.  In
contrast, little is known about entanglement in excited states, or how
this entanglement is related to quasiparticle interactions.  The
purpose of this paper is to address these latter issues by examining
entanglement in a model system of interacting qubits, the Heisenberg
spin chain.

The Heisenberg spin chain (HSC) is a one-dimensional lattice of
spin-$1/2$ particles, with nearest-neighbor spins coupled by the
exchange interaction.  The HSC Hamiltonian is

\begin{equation}\label{HSCHamiltonian}
    H_{HSC} = J \sum^{N}_{i=1} {\mathbf{S}}_{i} \boldsymbol{\cdot}
    {\mathbf{S}}_{i+1} ,
\end{equation}
where ${\mathbf{S}}_{i}=(S_{i}^{x},S_{i}^{y},S_{i}^{z})$ is the spin
operator associated with the spin at the $i$th site of the lattice;
its Cartesian components obey the usual angular momentum algebra. 
Periodic boundary conditions are assumed, so that
${\mathbf{S}}_{N+1}={\mathbf{S}}_{1}$.  $J$ is the exchange coupling
constant.  States in the $2^{N}$-dimensional Hilbert space
${\mathcal{H}}$ of the model can be expanded in the ``standard'' basis
of states of the form $\vert i,j, \ldots, k \rangle$, where the listed
spins $i,j,\ldots,k$ are aligned in the $+z$-direction (`up') and the
remaining spins are antiparallel (`down').  $\vert \emptyset \rangle$
denotes the state with all spins down.  The HSC and its anisotropic
relatives have been the subject of considerable previous research in
quantum information theory \cite{ArnesenBoseVedral},\cite{QPT
refs},\cite{thermal entanglement refs}.

The HSC model was originally solved by a method now termed the Bethe
Ansatz (BA) \cite{Bethe}.  In the physical interpretation of this
solution eigenstates are comprised of interacting quasiparticle
excitations called magnons; the $a$th magnon is characterized by a
pseudomomentum $k_{a}$, and there is a phase $\phi_{ab}$ between
magnons $a$ and $b$.  We would like to understand how the entanglement
between the spins of the HSC, viewed as qubits, depends on the
pseudomomenta and phases of the magnons.  Unfortunately, as we will
see, there is in general no simple relationship connecting qubit
entanglement to these parameters.  Nevertheless, the Bethe Ansatz
formalism provides information that is not available from a direct
numerical diagonalization of the Hamiltonian (although the BA
equations themselves often must be solved numerically), and this
information has much to tell us about the structure of qubit
entanglement in the spin chain.  Because the BA formalism is essential
to the discussion, it is reviewed very briefly in the following
section.

The entanglement between two spins will be quantified by calculating
their concurrence \cite{concurrence}, a measure of the inseparability
of two-qubit pure or mixed states commonly used in quantum information
theory.  Other metrics, such as the entanglement of formation, can be
calculated directly from the concurrence.  Because the HSC Hamiltonian
$H_{HSC}$ commutes with the total $z$-spin operator $S^{z} =
\sum_{i=1}^{N} S_{i}^{z}$, the reduced density matrix (RDM) for any
two spins in the states of interest always takes the form

\begin{equation}\label{RDM}
    \rho = \left( \begin{array}{cccc} \alpha & 0 & 0 & 0 \\ 
    0 & \beta & \gamma & 0 \\ 0 & \gamma^{*} & \delta
    & 0 \\ 0 & 0 & 0 & \epsilon \end{array} \right) .
\end{equation}
For such a density matrix, the concurrence is

\begin{equation}\label{concurrence}
    C = 2 \max \left( 0, \vert \gamma \vert - \sqrt{\alpha \epsilon}
    \right) .
\end{equation}
Concurrence ranges from zero, for a separable state, to one, for a
maximally entangled state.  Because the states of the HSC which will
be examined are translationally invariant, the concurrence depends
only on the separation between the two spins.  The concurrence between
nearest-neighbor spins will be denoted $C_{1}$, between
next-nearest-neighbor spins $C_{2}$, etc.

\section{Bethe Ansatz solution of the HSC}

The Bethe Ansatz consists of the hypothesis that the coefficients of
the expansions of the eigenstates of the HSC Hamiltonian
(Eq.~\ref{HSCHamiltonian}) with respect to the standard basis have the
form of a permutation-invariant sum of $n!$ exponential terms, where
$n$ is the number of spins inverted with respect to the reference
state $\vert \emptyset \rangle$ \cite{Bethe}.  For example, the
two-magnon eigenstates are assumed to have the form

\begin{eqnarray}\label{twomagnonstate}
    \lefteqn{\vert \psi (k_{1},k_{2},\phi_{12}) \rangle = \nu 
    (k_{1},k_{2},\phi_{12})
    \sum_{m_{1}<m_{2}}} \\
    && \left( e^{i k_{1} m_{1} + i k_{2} m_{2} +i \phi_{12}/2} + 
    e^{i k_{2} m_{1} + i k_{1} m_{2} - i \phi_{12}/2} \right)
    \vert m_{1},m_{2} \rangle , \nonumber
\end{eqnarray}
where $\nu$ is a normalization constant.  Substituting this state into
the Schr\"{o}dinger eigenvalue equation with the HSC Hamiltonian
yields the constraints (the Bethe Ansatz equations, or BAE) which
$k_{1}$, $k_{2}$, and $\phi_{12}$ must satisfy for the hypothesis to
be correct:

\begin{eqnarray} \label{2mBAE}
     2 \cot \left( \frac{\phi_{12}}{2} \right) & = & \cot \left( 
     \frac{k_{1}}{2} \right) - \cot \left( \frac{k_{2}}{2} \right) , \\
     N k_{1} - \phi_{12} & = & 2 \pi \lambda_{1} , \nonumber \\
     N k_{2} + \phi_{12} & = & 2 \pi \lambda_{2} . \nonumber
\end{eqnarray}
Note that under the interchange $k_{1} \leftrightarrow k_{2}$,
$\phi_{12}$ changes sign, so that the wavefunction coefficients in
Eq.~\ref{twomagnonstate} are permutation-invariant, as desired. 
$\lambda_{1}$ and $\lambda_{2}$ are integral quantum numbers between
$0$ and $N-1$ whose allowed values are prescribed by an arcane set of
rules \cite{Bethe}.  $k_{1}$ and $k_{2}$ are interpreted as the
pseudomomenta of two magnons, and $\phi_{12}$ is a phase (the Bethe
phase) between them.

The BA leads to a natural classification of two-magnon states as
either scattering states or bound states.  In scattering states each
magnon has a real pseudomomentum between $0$ and $2\pi$, and the Bethe
phase is real and between $0$ and $\pi$ if one chooses $k_{1} \leq
k_{2}$.  For such states the state vector (Eq.~\ref{twomagnonstate})
can be written in the form

\begin{eqnarray}\label{scatteringstate}
    \lefteqn{\vert \psi (k_{1},k_{2},\phi_{12}) \rangle = 2 \nu
    (k,\phi_{12}) \sum_{m_{1}<m_{2}} } \\ && e^{iK(m_{1}+m_{2})/2}
    \cos {\scriptstyle{\frac{1}{2}}} \left( k (m_{2} - m_{1}) +
    \phi_{12} \right) \vert m_{1},m_{2} \rangle , \nonumber
\end{eqnarray}
where $k \equiv k_{2}-k_{1}$ is the relative pseudomomentum, $K \equiv
k_{2}+k_{1}$ is the total pseudomomentum, and the normalization
constant $\nu (k,\phi_{12})$ is

\begin{eqnarray}\label{scatteringnormalization}
    \lefteqn{\nu^{-2}(k,\phi_{12}) = N(N-1) + } \nonumber \\ &&
    \frac{N \cos (k - \phi_{12}) - (N-1) \cos \phi_{12} - \cos (kN -
    \phi_{12})} {1 - \cos (k)} .
\end{eqnarray}
This expression remains correct in the limit $k \rightarrow 0$. 
Even when the BAE are not satisfied, states of the above form
(Eq.~\ref{scatteringstate}) are properly normalized physical states
(but not HSC eigenstates).

In bound states the magnons have complex conjugate pseudomomenta:
$k_{1,2} = u \pm iv$.  The imaginary component $v$ causes the
probability distribution for the separation of the two inverted spins
to be maximal when the spins are adjacent, and to decay exponentially
as the separation between them increases.  Due to this exponentially
tight binding of the inverted spins, bound states exhibit entanglement
behavior quite different from that of the scattering states, which (as
will be shown) is controlled entirely by the binding parameter $v$. 
The bound states are of two types, both of which can usefully be put
into forms independent of the Bethe phase.  The $\cosh$-type bound
(CB) states can be written as

\begin{eqnarray}\label{CBS}
     \vert \psi_{C} (u,v) \rangle & = & \nu_{C} (v)
     \sum_{m_{1}<m_{2}} e^{i u (m_{1} + m_{2})} \\ && \cosh \left[ v 
     \right( \frac{N}{2}-(m_{2} - m_{1}) \left) \right] \vert 
     m_{1},m_{2} \rangle , \nonumber
\end{eqnarray}
where

\begin{equation}\label{CBSnormalization}
    \nu_{C}(v) = \left[ \frac{4 \sinh v}{ N \left[
    \sinh \left( (N-1) v \right) + (N-1) \sinh v \right]}
    \right]^{\frac{1}{2}} ,
\end{equation}
while the $\sinh$-type bound (SB) states can be written as

\begin{eqnarray}\label{SBS}
     \vert \psi_{S} (u,v) \rangle & = & \nu_{S}(v)
     \sum_{m_{1}<m_{2}} e^{i u (m_{1} + m_{2})} \\ && \sinh
     \left[ v \right( \frac{N}{2}-(m_{2} - m_{1}) \left) \right] \vert
     m_{1},m_{2} \rangle , \nonumber
\end{eqnarray}
where

\begin{equation}\label{SBSnormalization}
    \nu_{S}(v) = \left[ \frac{4 \sinh v}{ N \left[
    \sinh \left( (N-1) v \right) - (N-1) \sinh v \right]}
    \right]^{\frac{1}{2}} .
\end{equation}

More generally, when there are $n$ magnons, the $n$ pseudomomenta and
$n(n-1)/2$ phases satisfy a set of $n$ linear and $n(n-1)/2$
transcendental coupled BA equations.  Multimagnon states can
consist of all scattering magnons with real pseudomomenta, but more
exotic states also appear.  These include ``wavecomplexes'' (to use
Bethe's terminology \cite{Bethe}), which are solitons in which a group
of magnons are all mutually bound, and mixtures of one or more
wavecomplexes with scattering magnons.  Note that instead of choosing
as the reference state $\vert \emptyset \rangle$, the state with all spins
down, one could equally well have chosen the state with all spins up,
and statements which hold for $n$-magnon states can therefore be
translated into equivalent statements for $(N-n)$-magnon states.

\section{Entanglement quenching}

\subsection{Two-magnon entanglement quenching}

The zero-magnon eigenstate of the HSC is the reference state $\vert
\emptyset \rangle$; it is of course a pure product state.  The excitation
of a single magnon changes this dramatically.  The one-magnon
eigenstates have the expected BA form

\begin{equation}\label{onemagnonstate}
       \vert \psi (k_{1}) \rangle = \frac{1}{\sqrt{N}}
       \sum_{m=1}^{N} e^{ik_{1}m} \vert m \rangle,
\end{equation}
where $k_{1}=2 \pi \lambda_{1} /N$ and $\lambda_{1}$ is a quantum
number that takes integral values between $0$ and $N-1$.  Remarkably,
the spins of the HSC are now \emph{equientangled}: the concurrence
between any two spins is $2/N$ irrespective of the pseudomomentum of
the magnon or of the separation of the spins on the chain.

What is the effect on entanglement of the excitation of a second
magnon?  To begin, we consider the special case where the second
magnon has zero pseudomomentum.  Because of the magnon dispersion
relation $E_{k} \sim 1-\cos (k)$, no energy is required to excite such
a magnon: it is ``massless''.  Such modes can be considered as arising
from the spontaneous breaking of the global SU(2) symmetry of the HSC
Hamiltonian (Eq.  \ref{HSCHamiltonian}) by the choice of the reference
state $\vert \emptyset \rangle$, and hence will be referred to as Goldstone
magnons \cite{GoldstoneSalamWeinberg}.

From the two-magnon BA equations (Eq.~\ref{2mBAE}), we see that when
one of the pseudomomenta vanishes, the Bethe phase $\phi_{12}$ must
also vanish.  The HSC is now in the state

\begin{eqnarray}\label{GM+nGM}
    \lefteqn{\vert \psi \left( k_{1}=0, k_{2}, \phi_{12}=0 \right)
    \rangle = 2 \nu (k_{2},0) \sum_{m_{1}<m_{2}} } \nonumber \\ && e^{i
    k_{2} (m_{1}+m_{2})/2} \cos \left( k_{2} (m_{2} -
    m_{1}) /2 \right) \vert m_{1},m_{2} \rangle .
\end{eqnarray}
We can visualize the entanglement in this state by plotting the
concurrence between two spins as a function of the relative
pseudomomentum $k=k_{2}$, treating $k$ as a continuous parameter, as
in Fig.~\ref{N=6,phi=0}.  These states are eigenstates precisely when
$k=2 \pi \lambda_{2} /N$ ($\lambda_{2}=0,\ldots,N-1$), which is also
the condition for translational invariance, and so only the separation
of the spins matters at the points of interest.  When $k=0$, all qubit
concurrences are equal with $C_{i} \approx 0.21$.  Strikingly, all
other eigenstates correspond to exact zeros of the concurrences
$C_{i}$.  The excitation of a Goldstone magnon has completely quenched
the qubit equientanglement, except in the special case in which the
original magnon had zero pseudomomentum as well.

\begin{figure}
\begin{center}
     \includegraphics{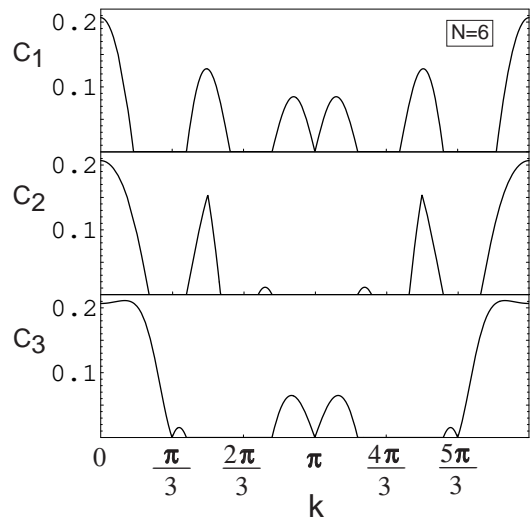}
\end{center}
\caption{Concurrence between spins 1 and 2 (\emph{top}), 1 and 3
(\emph{middle}), and 1 and 4 (\emph{bottom}) in a two-magnon
scattering state $\vert \psi (0,k,0) \rangle$ of the six-spin
Heisenberg chain.  The relative pseudomomentum $k$ is treated as a
continuous variable parameterizing states of the scattering form
(Eq.~\ref{scatteringstate}); $\phi_{12}$ is set to zero.  Eigenstates
occur at $k=2\pi \lambda_{2} /N$, $\lambda_{2}=0,\ldots,5$, and are
indicated by ticks on the $k$-axis.}
\label{N=6,phi=0}
\end{figure}

It is possible to prove analytically that this must always occur
whenever a Goldstone and a non-Goldstone magnon mix.  Choose two
arbitrary spins $p<q$ of the HSC and trace out the rest: this yields
the reduced density matrix $\rho (p,q)$, whose elements are (in the
notation of Eq.~\ref{RDM}):

\begin{eqnarray}
    \alpha (p,q) & = & \frac{4}{N(N-2)} c^{2}(\lambda_{2}) \\
    \beta  (p,q) & = &  \delta (p,q) = \frac{2}{N(N-2)} \left( N-2-2 c^{2}(\lambda_{2}) \right)  \\
    \gamma (p,q) & = & \frac{2(N-4)}{N(N-2)} \exp \left( \pi i 
    (p-q) \lambda_{2} / N \right) c(\lambda_{2}) \\
    \epsilon (p,q) & = & \frac{1}{N(N-2)} \left( (N-2)(N-4) + 4 c^{2}(\lambda_{2}) 
    \right) ,
\end{eqnarray}
where

\begin{equation}
    c(\lambda_{2}) = \cos \left( \frac{\pi \lambda_{2} (p-q)}{N} 
    \right) ,
\end{equation}
and $\lambda_{2} \in \{1,\ldots,N-1 \}$ (these matrix elements are
\emph{not} correct when $\lambda_{2}=0$).  The concurrence between
spins $p$ and $q$, from Eq.~\ref{concurrence}, is

\begin{eqnarray}
    \lefteqn{C(p,q) = \frac{4 \vert c(\lambda_{2}) \vert}{N(N-2)} 
    \times } \nonumber 
    \\ && \max \left( 0, N-4-\sqrt{(N-2)(N-4)+4 c^{2}(\lambda_{2})} 
    \right) .
\end{eqnarray}
But

\begin{eqnarray}
    \sqrt{(N-2)(N-4)+4 c^{2}(\lambda_{2})} & > & \sqrt{(N-2)(N-4)} 
    \nonumber \\ 
    & > & N-4
\end{eqnarray}
for $N \geq 4$ and hence $C(p,q)$ is identically zero.

Because the two-magnon state under consideration (Eq.~\ref{GM+nGM}) is
pure, the entanglement of formation $S_{f}$ for the reduced density
matrix $\rho (p,q,r,\ldots)$ of any subset $p,q,r,\ldots$ of the spins
of the HSC coincides with the binary von Neumann entropy of that
density matrix: $S_{f} = - \textrm{Tr} \left[ \rho (p,q,r,\ldots)
\textrm{lg} \rho (p,q,r,\ldots) \right]$, where ${\textrm{lg}}(z) =
\log_{2} (z)$.  In particular, the entanglement of formation for the
state of a single spin $p$ is simply

\begin{equation}
	    S_{f} = - \left( \frac{2}{N} \right) \textrm{lg} \left(
	    \frac{2}{N} \right) - \left( 1-\frac{2}{N} \right)
	    \textrm{lg} \left( 1-\frac{2}{N} \right) .
\end{equation}
This is nonzero for any finite $N>2$.  These results yield the somewhat
paradoxical conclusion that in states containing both a Goldstone and
a non-Goldstone magnon, any specific spin is not entangled with any
other spin individually, although it is entangled with all other spins
collectively.

\subsection{Multimagnon entanglement quenching}

Suppose that a third magnon is excited on a HSC initially in a
two-magnon entanglement-quenched state (Eq.~\ref{GM+nGM}).  Will this
create qubit entanglement?  Again we can begin with the special case
in which the additional magnon is also a Goldstone excitation. 
Three-magnon state vectors are somewhat difficult to manipulate
analytically.  They can however be generated with arbitrary precision
using a numerical implementation of the Bethe Ansatz, and the partial
trace required to find the two-spin RDM and the concurrence can then
be done numerically with no loss of accuracy.  All possible
concurrences in all states describing the simultaneous excitation of
two Goldstone and one non-Goldstone magnons for Heisenberg spin chains
of lengths from $N=6$ to $N=50$ were calculated in this manner and
found to be zero.  Thus a second Goldstone magnon cannot revive
quenched entanglement.

A different situation arises when two non-Goldstone magnons are
present.  Such states have complex entanglement structures, as will be
discussed later.  Table~\ref{N=6concurrences} shows the effects of the
excitation of a Goldstone mode on qubit entanglement in the two-magnon
scattering and bound states of an $N=6$ ring (in this case the BAE can
be handled analytically).  In five of the states entanglement is
completely quenched.  In two of the bound states (quantum numbers
$\lambda_{1}=\lambda_{2}=1 \ \textrm{or} \ 5$) the qubit entanglement
is reduced, but not eliminated, while in two of the scatting states
($\lambda_{1}=1$, $\lambda_{2}=5$ and $\lambda_{1}=2$,
$\lambda_{2}=4$), entanglement is actually generated at the longest
range (i.e. $C_{3}$).  Nevertheless, the \emph{total} entanglement, as
measured by the sum of the concurrences $C_{i}$, has decreased.  This
phenomenon is confirmed by numerical studies of other rings:
excitation of a Goldstone magnon will occasionally generate
entanglement between a formerly unentangled pair, but the total
entanglement of the HSC is always reduced.

Finally, we consider entanglement quenching in the four-magnon case. 
Suppose a second Goldstone magnon is excited on a ring with two
non-Goldstone and one Goldstone magnons.  Numerical examination of all
pairs of spins for Heisenberg spin chains of lengths from $N=8$ to
$N=20$ show that all concurrences vanish in such states.  Concurrence
is also absent from these chain in all states with three Goldstone and
one non-Goldstone magnons.  These results, and those of the preceeding
section, support the following hypothesis: the addition of a Goldstone
excitation to a state containing one or more non-Goldstone magnons
always reduces the total qubit entanglement in the HSC; and when the
number of Goldstone magnons equals or exceeds the number of
non-Goldstone excitations, there is no qubit entanglement whatsoever.

\begin{table}[b]
    \begin{center}
	\begin{ruledtabular}
	\begin{tabular}{cc|ccc|ccc}
	    $\lambda_{1}$ & $\lambda_{2}$ &
	    $C_{1}$ & $C_{2}$ & $C_{3}$ &
	    $C_{1}$ & $C_{2}$ & $C_{3}$ \\ \hline
	    1 & 3 & 0.45 & 0 & 0 & 0 & 0 & 0 \\
	    1 & 4 & 0.04 & 0 & 0 & 0 & 0 & 0 \\
	    1 & 5 & 0 & 0.10 & 0 & 0 & 0 & 0.06 \\
	    2 & 4 & 0.43 & 0 & 0 & 0.21 & 0 & 0.06 \\
	    2 & 5 & 0.04 & 0 & 0 & 0 & 0 & 0 \\
	    3 & 5 & 0.45 & 0 & 0 & 0 & 0 & 0 \\ \hline
	    1 & 1 & 0 & 0.28 & 0.35 & 0 & 0 & 0.15 \\
	    4 & 5 & 0 & 0.34 & 0 & 0 & 0 & 0 \\
	    5 & 5 & 0 & 0.28 & 0.35 & 0 & 0 & 0.15 \\
	\end{tabular}
        \end{ruledtabular}
    \end{center}
\caption{Partial quenching of entanglement.  $\lambda_{1}$ and
$\lambda_{2}$ are the quantum numbers of two non-Goldstone magnons on
a six-spin HSC. Concurrences $C_{i}$ are shown without (\emph{center
column}) and with (\emph{right column}) an additional Goldstone
magnon.  The first six entries are scattering states, while the next
three are bound states.}
\label{N=6concurrences}
\end{table}

\section{The pure Goldstone sector}

\begin{figure}
\begin{center}
     \includegraphics{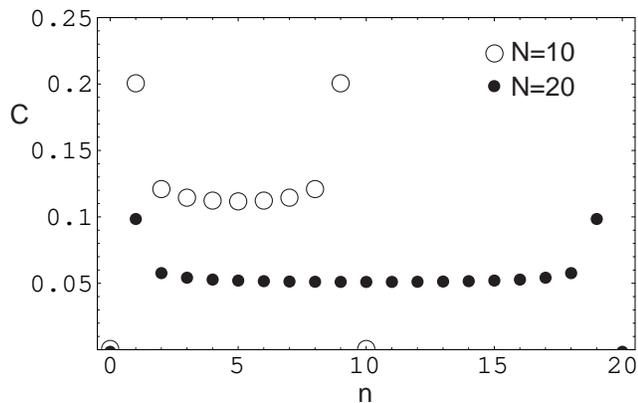}
\end{center}
\caption{Concurrence between any two qubits in a $N$-spin Heisenberg
chain as a function of the number $n$ of Goldstone magnons present.}
\label{GoldstoneSector}
\end{figure}

As remarked previously and illustrated in Table.~\ref{N=6,phi=0}, the
excitation of second Goldstone magnon from a single-Goldstone state
reduces but does not completely quench qubit entanglement.  Thus from
the standpoint of entanglement behavior, the pure Goldstone sector is
fundamentally different from the mixed sector.  As further Goldstone
magnons are excited, the qubit concurrence continues to decrease
without vanishing.  There is no length scale associated with these
Goldstone magnons (because their pseudomomentum is zero), and the HSC
qubits remain equientangled in all states comprised entirely of
Goldstone excitations.  This situation can be studied analytically. 
The general formula for the concurrence between any two qubits in an
$n$-Goldstone state of an $N$-qubit ring is

\begin{eqnarray}
    \lefteqn{C(n,N) = \frac{2}{N(N-1)} \times} \nonumber \\ && [ n(N-n) -
    \sqrt{n(n-1)(N-n)(N-n-1)} ] .
\end{eqnarray}
This result has been derived elsewhere, and its implications for
thermal entanglement have been discussed \cite{Pratt}.  When $n=0$ or
$n=N$, the eigenstate corresponds to all spins pointing down (the
state $\vert \emptyset \rangle$) or all spins pointing up, and hence is
factorable.  Maximal entanglement occurs when a single magnon is
excited with respect to either of these states, i.e. $n=1$ or $n=N-1$. 
Excitation of a second Goldstone magnon produces a large relative
decline in entanglement, but thereafter qubit entanglement is only
weakly dependent on the number of Goldstone magnons excited, as shown
in Fig.~\ref{GoldstoneSector}.

\begin{figure}[b]
\caption{\emph{(right)} Contour plots of concurrences between spins 1
and 2 ($C_{1}$), spins 1 and 3 ($C_{2}$), etc., for two-magnon
scattering states (Eq.~\ref{scatteringstate}) of an 8-spin HSC, as
functions of the relative pseudomomentum $k$ and the Bethe phase
$\phi=\phi_{12}$.  White regions denote zero concurrence; contours are
evenly spaced at intervals of 0.05 (\emph{top two graphs}), 0.03
(\emph{third graph}), and 0.01 (\emph{bottom graph}).  Values of
$(k,\phi)$ corresponding to energy eigenstates are indicated by dots. 
Small dots denote a single eigenstate.  The six large dots denote two
eigenstates with the same $(k,\phi)$ values and hence the same
entanglement; such a pair need not be energetically degenerate.  (Note
that near $k \approx 1,\phi \approx 2.5$ a single state and a double
state nearly coincide.)  The point at $(0,0)$ is the eigenstate
containing two Goldstone magnons.  The seven eigenstates with one
Goldstone magnon form a row on the $\phi$--axis.  In the $C_{4}$ plot
the dotted lines (for example, bisecting the central ``islands'') are
zero-entanglement boundaries between two entangled regions.}
\label{ScatteringCon}
\end{figure}

\begin{figure}
\begin{center}
     \includegraphics{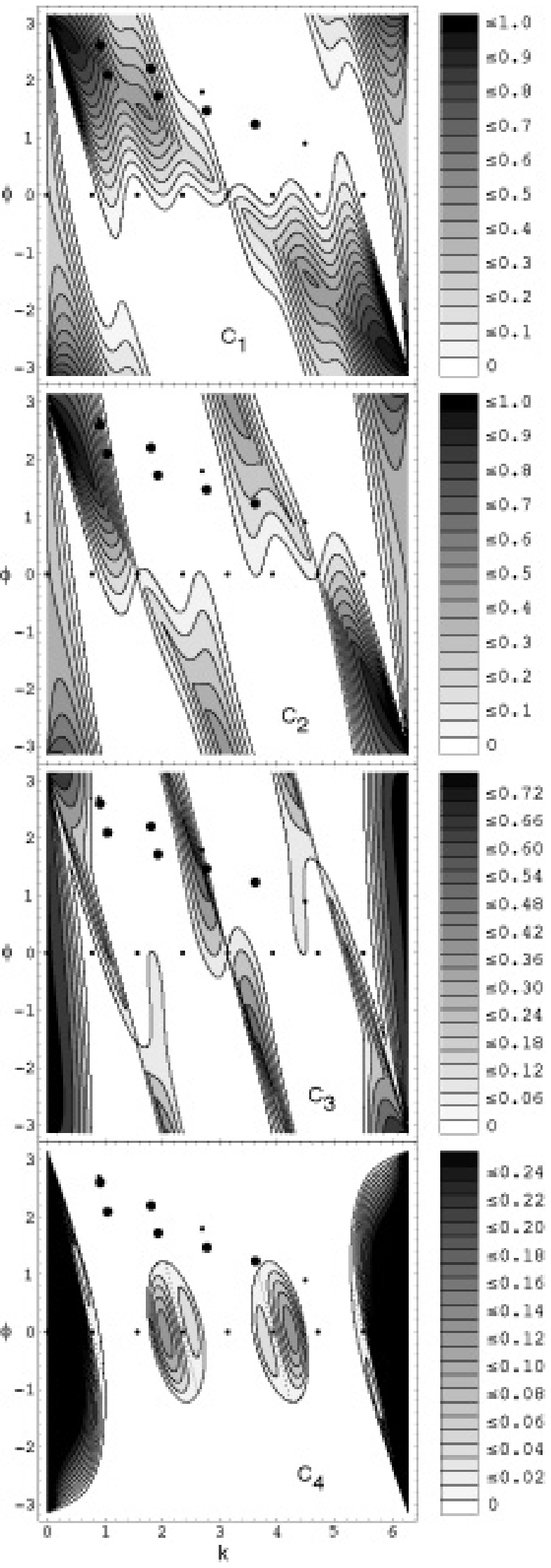}
\end{center}
\end{figure}

\begin{figure*}[t]
\begin{center}
     \includegraphics{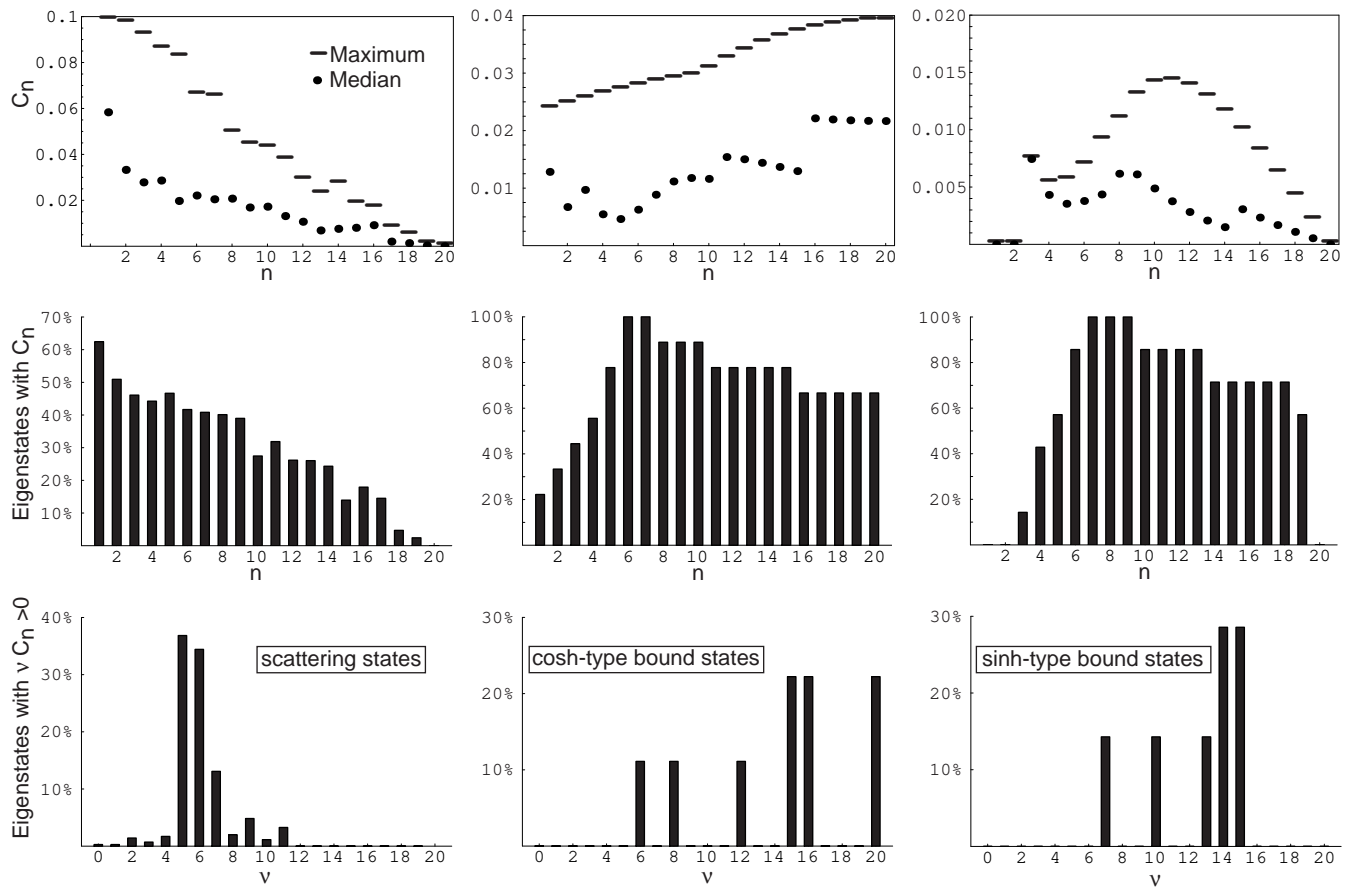}
\end{center}
\caption{Statistical characterization of entanglement in two-magnon
states for the $N=40$ HSC. \emph{Left column}: scattering eigenstates. 
\emph{Center column}: $\cosh$-type bound eigenstates.  \emph{Right
column}: $\sinh$-type bound eigenstates.  The upper graphs show the
maximum concurrence at a qubit separation $n$ among all eigenstates in
the indicated population, and the median concurrence of the
subpopulation of states with nonzero $C_{n}$.  The middle graphs show
the percentage of two-magnon eigenstates with nonzero $C_{n}$ as a
function of $n$.  The lower graphs show the percentage of two-magnon
scattering states states with exactly $\nu$ values of $C_{n}$ not
equal to zero.}
\label{StateStatistics}
\end{figure*}

\section{The Goldstone-free sector}

\subsection{Two-magnon scattering states}

The previous sections have characterized qubit entanglement in the
pure Goldstone and the mixed Goldstone/non-Goldstone sectors of the
HSC. We now consider the pure non-Goldstone sector, beginning with the
two-magnon states.  For scattering states (Eq.~\ref{scatteringstate})
it is easy to show that the concurrence depends only on the relative
pseudomomentum $k$ and the Bethe phase $\phi$, and not on the total
pseudomomentum $K$.  We can plot the concurrence as a function of
these variables unconstrained by the BAE (Eq.~\ref{2mBAE}).  This is
shown in Fig.~\ref{ScatteringCon} for the $N=8$ case, with the
eigenstates (those parameter values which do satisfy the BAE)
indicated by dots.  The fifteen zero-Goldstone states form an
irregular wedge in the upper half-plane.  The double-Goldstone state
and the single-Goldstone states appear as a row on the $\phi$--axis. 
As in Fig.~\ref{N=6,phi=0}, the positioning of these latter states is
striking: although always unentangled, they often lie near or on the
boundaries of regions of entanglement, so that a slight perturbation
of the Bethe parameters would yield an entangled state.  There is no
relation between an eigenstate's entanglement and its eigenenergy
(which, unlike its concurrences, depends on the total pseudomomentum),
and, as the complexity of the concurrence diagrams suggests, there is
no simple relationship between entanglement and the Bethe parameters.

Although qubit entanglement cannot easily be calculated directly in
terms of the Bethe parameters for non-Goldstone multimagnon states,
the Bethe Ansatz classification partitions eigenstates into
populations with statistically distinct patterns of qubit
entanglement.  In the two-magnon case, the scattering states are
characterized qualitatively by short-range entanglement, which
distinguishes them from the bound states.  There are a number of ways
to make this observation quantitative.  One way is to ask about the
maximum concurrence attained by \emph{any} scattering state as a
function of qubit separation.  As shown in Fig.~\ref{StateStatistics}
for the $N=40$ case, this maximum decreases roughy linearly (but
nonmonotonically) with increasing qubit separation, and in fact is
exactly zero at the longest range ($C_{20}$), that is, between
opposite spins on the Heisenberg chain.  This is not simply due to
outlying extreme values.  If one culls the total population of
two-magnon scattering states, and calculates the median concurrence as
a function of qubit separation $n$ among those states with nonzero
$C_{n}$, the median of each subpopulation is typically between
one-half and one-third of the maximum.  Another measure of the
short-range nature of entanglement is the decline in the number of
eigenstates with nonzero concurrence as a function of distance.  Over
$60\%$ of scattering states have nearest-neighbor entanglement; this
decays roughly linearly with increasing qubit separation, and as
already noted becomes zero at the longest range.  Most two-magnon
scattering states have entanglement at only a few different distances,
usually only $5$ or $6$; no scattering state has more than $11$
nonzero values of $C_{n}$.  Similar observations hold for other values
of $N$.

Examination of $C_{4}$ entanglement in Fig.~\ref{ScatteringCon} (where
$N=8$) shows that all the scattering states lie in regions of zero
concurrence.  Similarly, inspection of Table~\ref{N=6concurrences}
shows that $C_{3}$ vanishes for scattering states when $N=6$. 
Analytic calculations showed that $C_{2}$ was always zero for
scattering states when $N=4$ or $5$, and $C_{20}$ was seen above to
vanish when $N=40$.  These observations suggested the hypothesis that
concurrence at the longest range, $C_{\lfloor N/2 \rfloor}$, is always
zero for two-magnon scattering states.  All such states for $N=4$ up
to $N=50$ were generated numerically, and $C_{\lfloor N/2 \rfloor}$
always equalled zero as conjectured.  This is yet another
manifestation of the short-range nature of scattering state
entanglement.

\subsection{Two-magnon bound states}

\begin{figure}[t]
\begin{center}
     \includegraphics{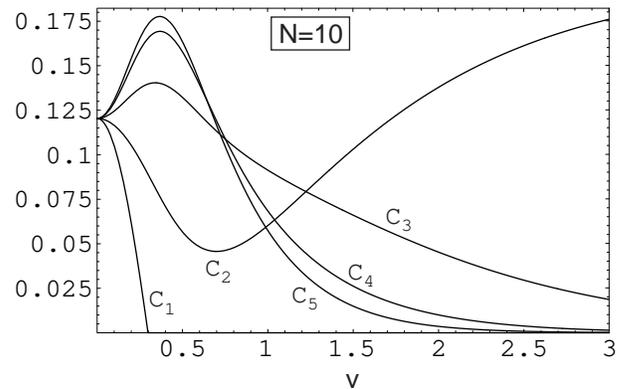}
\end{center}
\caption{Nearest-neighbor concurrence $C_{1}$, next-nearest-neighbor
concurrence $C_{2}$, etc., for $\cosh$-type bound (CB) states in the
10-spin Heisenberg chain.  Two eigenstates occur at $v=0.258$ and two
at $v=1.174$.  An `anomalous' regime, in which entanglement increases
with qubit separation, extends to about $v \approx 0.7$.}
\label{CoshBoundCon}
\end{figure}

\begin{figure}[t]
\begin{center}
     \includegraphics{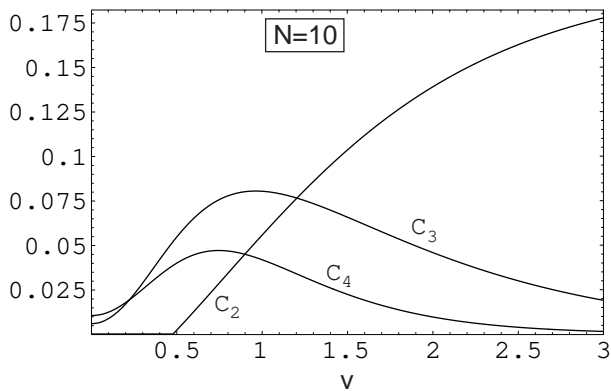}
\end{center}
\caption{Next-nearest-neighbor concurrence $C_{2}$, etc., for
$\sinh$-type bound (SB) states in the 10-spin Heisenberg chain.  Two
eigenstates occurs at $v=0.521$; the singular state
(Eq.~\ref{singularstate}) lies at $v=\infty$.  Note the absence of
both nearest-neighbor ($C_{1}$) and longest-range ($C_{5}$)
entanglement.}
\label{SinhBoundCon}
\end{figure}

In contrast to the scattering states, the two-magnon bound states are
characterized by long-range entanglement.  The behavior of the
$\cosh$-type (Eq.~\ref{CBS}) and $\sinh$-type (Eq.~\ref{SBS}) states
is similar but distinct.  Analogous to the independence of scattering
state entanglement from the total pseudomomentum $K$, bound state
entanglement is easily shown to be independent of the phase parameter
$u$; qubit concurrence is controlled entirely by the binding parameter
$v$.  This dependence can be plotted by treating $v$ as a continuous
parameter unconstrained by the BAE (Eq.~\ref{2mBAE}), as shown in
Fig.~\ref{CoshBoundCon} for the CB states and in
Fig.~\ref{SinhBoundCon} for the SB states for a 10-spin HSC. In the
limit $v \rightarrow 0$, the CB state becomes the two-Goldstone state
and all concurrences are equal.  As $v$ increases, the short-range
components of entanglement, $C_{1}$ and $C_{2}$, fall off, while the
longer-range components increase.  In this low-$v$ `anomalous'
parameter regime $C_{n}$ increases with $n$, that is, spins actually
become more highly entangled with increasing separation.  As $v$
continues to increase, $C_{1}$ becomes identically zero, and never
revives.  $C_{2}$ passes through a minimum and rises to become the
dominant component, and the order of the strengths of the longer-range
concurrences undergoes an inversion, so that entanglement now falls
off with increasing spin separation.  Asymptotically, as $v
\rightarrow \infty$, all concurrences vanish except for $C_{2}$, whose
limit will be calculated below.  Similar features occur for longer-$N$
spin chains; in particular, there is always an `anomalous' region for
small $v$.

The physical implications of this for eigenstates are determined by
the distribution of solutions to the BAE (Eq.\ref{2mBAE}), which for
the CB states can be rewritten as

\begin{equation}\label{coshBAE}
    \coth \left( \frac{Nv}{2} \right) = \frac{\sinh v}{\cosh v - \cos 
    \left( \frac{\pi \lambda}{N} \right)},
\end{equation}
where $\lambda$ is a quantum number.  Numerically, it is found that
the roots (values of $v$ corresponding to eigenstates) of this
equation always lie between $N^{-3/2}$ (a strict lower bound) and
$\ln(N)$ (a strict upper bound), with one exception which will be
described below; further, the roots are strongly clumped towards the
lower bound.  The result is that most allowed $v$-values actually lie
in the anomalous parameter region where entanglement strength
increases with qubit separation; this becomes more pronounced as $N$
increases.  This is reflected in the statistics of the CB state
population, as shown in Fig.~\ref{StateStatistics}.  In general, CB
states are much more entangled than scattering states; for example,
nearly a quarter of the CB states have nonzero concurrence at every
length scale.  The magnitude of the concurrences, however, is
typically smaller for bound states than for scattering states.

When $N>2$ is even, but not when $N$ is odd, there is one solution of
the two-magnon BAE in which the pseudomomenta have infinite imaginary
parts \cite{Siddharthan}.  This is the exception mentioned above,
corresponding to $v \rightarrow \infty$.  This eigenstate with
singular Bethe parameters is:

\begin{equation}\label{singularstate}
    \vert \psi_{\infty} \rangle = \frac{1}{\sqrt{N}} 
    \sum_{m=1}^{N} (-1)^{m+1} \vert m,m+1 \rangle .
\end{equation}
It is the maximally bound state; the two inverted spins are always
adjacent.  Formally, it is a CB state when $N/2$ is even and an SB
otherwise.  When all spins except $p$ and $q$ are traced out, the
elements of the RDM are

\begin{eqnarray}\label{DM:alt}
    \alpha (p,q) & = & \frac{1}{N} \delta_{p+1,q} \\
    \beta (p,q) & = & \delta (p,q) = \frac{1}{N} (2 - \delta_{p+1,q}) \\
    \gamma (p,q) & = & \frac{-\delta_{p+2,q}-\delta_{N,4}}{N} \\
    \epsilon (p,q) & = & \frac{N - 4 + \delta_{p+1,q}}{N},
\end{eqnarray}
where $\delta_{ij}$ is the Kronecker delta.  The concurrence is
therefore

\begin{equation}\label{con:alt}
    C_{r} = \left( \frac{2+2\delta_{N,4}}{N} \right) \delta_{r,2} .
\end{equation}
Thus there is an unusual entanglement pattern, with no entanglement
except between next-nearest-neighbors, whose concurrence is simply
$2/N$ if \mbox{$N>4$}.  The four-spin ring is exceptional; opposite
corner qubits are perfectly entangled.  Intuitively, diagonal qubits
are next-nearest-neighbors on both sides (in both directions around
the ring).  It is striking that the $v \rightarrow \infty$ limit is
realized by a physical eigenstate.

There are several notable differences that distinguish the CB and SB
bound states.  First, in SB states nearest-neighbor entanglement is
always identically zero.  This can be proven analytically, by a method
resembling that used for the proof of Goldstone-mediated entanglement
quenching given above, but computationally much more involved. 
Second, when $N$ is even, longest-range entanglement $C_{N/2}$ is
also identically zero.  The result follows immediately from
calculating that the coherence $\gamma$ (notation of Eq.~\ref{RDM})
vanishes for opposite spins, irrespective of the value of $v$.  Thus
SB state entanglement exhibits a sensitivity to the parity of $N$ that
CB states do not.  Most significantly, however, the anomalous
parameter regime does not in general exist for SB states; rather, at
$v=0$, the maximum concurrence typically lies at some intermediate
range, so that $C_{N/4}$ (or thereabouts) is the dominant form of
entanglement.  As $v$ increases, the concurrence curves cross at
irregular intervals.  The values of $v$ corresponding to eigenstates
are constrained by the BAE:

\begin{equation}\label{sinhBAE}
    \tanh \left( \frac{Nv}{2} \right) = \frac{\sinh v}{\cosh v - \cos 
    \left( \frac{\pi \lambda}{N} \right)} .
\end{equation}
As with Eq.~\ref{coshBAE}, the roots cluster strongly at small $v$
values, and the eigenstate entanglement statistics
(Fig.~\ref{StateStatistics}, right column) reflect the dominance of
intermediate-range entanglement.  Thus the various classes of
two-magnon eigenstates as determined by the Bethe Ansatz exhibit
distinct patterns of entanglement behavior.

\subsection{Multimagnon states}

\begin{figure}[t]
\begin{center}
     \includegraphics{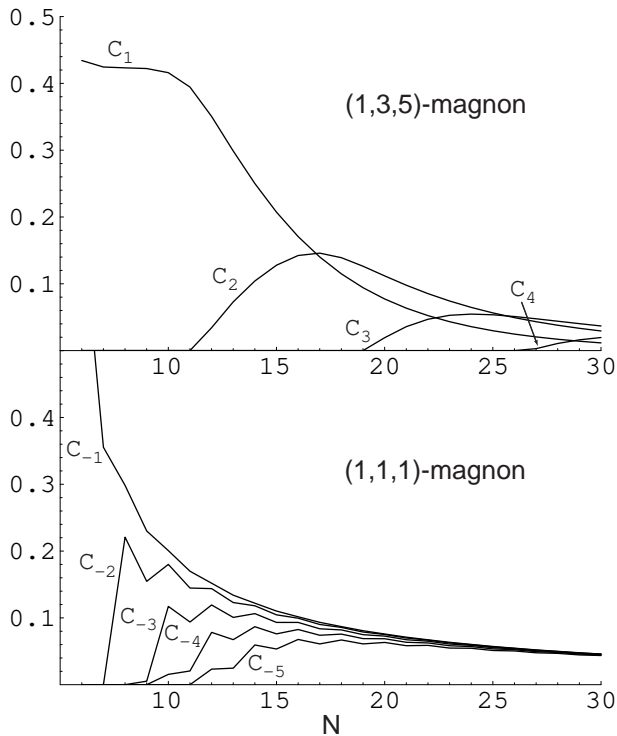}
\end{center}
\caption{Qubit concurrences in the three-magnon pure scattering
eigenstate with quantum numbers $\vec{\lambda}=(1,3,5)$, and in the
three-magnon pure bound eigenstate with quantum numbers
$\vec{\lambda}=(1,1,1)$, as functions of the HSC length $N$.  Curves
are drawn continuously for visual clarity, but only values at integral
$N$ are meaningful.  The notation $C_{-i}$ is used to denote the
concurrence $C_{\lfloor N/2 \rfloor +1-i}$, that is, $C_{-1}$ is the
concurrence at the longest range (between opposite qubits on the
ring), $C_{-2}$ is the concurrence at the next-longest range, etc. 
The (1,1,1)-magnon has shorter-range concurrences (lower $C_{-i}$
components) which are not shown.}
\label{(1,3,5)&(1,1,1)}
\end{figure}

The entanglement characteristics of two-magnon bound and scattering
states generalize to multimagnon states.  For example, the quantum
numbers $\vec{\lambda}=(\lambda_{1},\lambda_{2},\lambda_{3})=(1,3,5)$
specify a pure scattering state of three magnons with real
pseudomomenta for any chain length $N \geq 6$, while
$\vec{\lambda}=(1,1,1)$ always corresponds to a pure bound state, i.e.
a three-magnon wavecomplex.  The entanglement behaviors of these
eigenstates are shown in Fig.~\ref{(1,3,5)&(1,1,1)}.  The scattering
state has only nearest-neighbor entanglement up to $N=12$;
next-nearest-neighbor entanglement does not become dominant until
$N=17$, and $C_{3}$ entanglement arises only at $N=20$.  Thus,
qualitatively, the range of nonzero entanglement is always much
smaller than the longest range possible.  This behavior is typical of
all pure scattering multimagnon states studied so far.  In contrast,
the wavecomplex state initially ($N=6$) has only longest range
entanglement (written as $C_{-1}$, as explained in the caption for
Fig.~\ref{(1,3,5)&(1,1,1)}); next-longest range entanglement $C_{-2}$
arises at $N=8$, and by $N=16$ entanglement is present at every range. 
However, the magnitude of the concurrence always increases as qubit
separation increases, just as for a two-magnon CB state in the
`anomalous' regime.  Thus the entanglement patterns observed in the
two-magnon case also occur in multimagnon bound and scattering states.

Of particular interest is the behavior of the antiferromagnetic ground
state (AGS), which for even-$N$ chains is nondegenerate.  In the BA
formalism the AGS is specified by the quantum numbers
$\vec{\lambda}=(1,3,\ldots,N-1)$ and is a pure scattering state.  It
has been proven by O'Connor and Wootters \cite{O'ConnorWootters} that
of all even-$N$ translationally invariant states with $S^{z}=0$, the
AGS has maximal nearest-neighbor concurrence.  To study this further,
the AGS was calculated analytically for $N=4,6$ and numerically
for $N=8,10,12$ (work is underway to study longer HSCs).  Two new
results were obtained.  First, all concurrences except $C_{1}$
vanished; that is, the AGS appears (based on these few cases) to have
\emph{only} nearest-neighbor entanglement, as one might anticipate for
a pure scattering state with the maximum number of magnons possible. 
More generally, no examples have been found of a chain with $\lfloor
N/2 \rfloor$ scattering magnons having any qubit concurrence beyond
nearest-neighbor, although this work is preliminary.  Second, while
the AGS does indeed have the largest nearest-neighbor entanglement,
other translationally invariant $S^{z}=0$ eigenstates can have much
stronger entanglement at longer ranges.  For example, the 6-spin AGS
$\vec{\lambda}=(1,3,5)$ has $C_{1}=0.43456$, but the bound state
$\vec{\lambda}=(1,1,1)$ has $C_{3}=0.70313$.

\section{Discussion}

The Bethe Ansatz provides a method for describing eigenstates of the
Heisenberg spin chain in terms of their pseudoparticle content.  This
paper has shown that the natural classification of eigenstates
originating in this description predicts qualitatively and sometimes
quantitatively the qubit entanglement of these states.  Magnons with
zero pseudomomentum, here termed Goldstone magnons, effectively reduce
or suppress entanglement relative to the corresponding Goldstone-free
state.  From an applied point of view this is a form of noise which
may affect the processing, storage or coherent transmission of quantum
information in exchange-coupled qubits \cite{PrattEberly}.  Pure
scattering states are composed of one or more magnons with real
pseudomomenta.  While a single such magnon leads to qubit
equientanglement, the presence of additional scattering magnons favors
short-range entanglement: statistically, qubit entanglement becomes
both weaker and less common in the population of scattering states as
interqubit separation increases.  In particular, the antiferromagnetic
ground state seems to exhibit only nearest-neighbor entanglement.  In
contrast, bound states consist of two or more magnons with complex
pseudomomenta.  For a pair of bound magnons qubit entanglement is
controlled solely by the binding parameter $v$ and the interqubit
separation.  In the large-$v$ limit next-nearest-neighbor entanglement
approaches $2/N$, while all other forms of entanglement approach zero. 
This $v \rightarrow \infty$ limit is actually realized in even-$N$
spin chains by the singular state.  Most nonsingular two-magnon bound
states have small values of $v$; in this binding parameter regime the
order of the concurrence curves is inverted for $\cosh$-type states,
so that entanglement \emph{increases} with qubit separation, while in
$\sinh$-type bound states entanglement at intermediate lengths is
favored statistically.  Behavior similar to that of the $\cosh$-type
states is seen in multimagnon bound states such as the three-magnon
$\vec{\lambda}=(1,1,1)$ state.  Multi-wavecomplex states, comprised of
magnons bound into groups which are not bound to one another (e.g. a
four-magnon state of two bound pairs), remain to be studied, as do
mixtures of scattering magnons and wavecomplexes.

The outstanding question at this point is whether this approach to
understanding entanglement via the Bethe Ansatz can be extended to
other models.  The BA can also be used to solve other spin-$1/2$
chains, such as the anisotropic deformations of the HSC (e.g. the XXZ
and XYZ models), and it will be interesting to see if the same
features, such as short-range entanglement in scattering states,
reappear.  More challenging is the extension of this approach to
Hamiltonians which include hopping, such as the one-dimensional
Hubbard model, which are solved by the so-called nested Bethe Ansatz
\cite{LiebWu}.  If similar features do reappear, as seems plausible,
it may reflect a kind of universality arising from the BA structure of
the solutions, allowing us to predict aspects of entanglement behavior
across a broad range of physical systems.

\end{document}